\documentclass[%
reprint,
superscriptaddress,
amsmath,amssymb,
aps,
]{revtex4-1}

\usepackage{color}
\usepackage{graphicx}
\usepackage{dcolumn}
\usepackage{bm}

\begin {document}

\title{Statistics of the number of renewals, occupation times and correlation in ordinary, equilibrium and aging  alternating renewal processes}

\author{Takuma Akimoto}
\email{takuma@rs.tus.ac.jp}
\affiliation{%
  Department of Physics, Tokyo University of Science, Noda, Chiba 278-8510, Japan
}%


\date{\today}

\begin{abstract}
Renewal process is a point process where an inter-event time between successive renewals 
is an independent and identically distributed random variable. Alternating renewal process is a dichotomous process and 
 a slight generalization of the renewal process, where the inter-event time distribution alternates between two distributions. 
We investigate statistical properties of the number of renewals and occupation 
times for one of the two states in alternating renewal processes. When both means of the inter-event times are finite, 
the alternating renewal process can reach an equilibrium. On the other hand, an alternating renewal process 
shows aging when one of the means diverges. We provide analytical calculations for the moments of the number of renewals, occupation time statistics, and 
the correlation function for several case studies in the inter-event-time distributions. We show anomalous fluctuations for the number of renewals and occupation times 
when the second moment of inter-event time diverges.
When the mean inter-event time diverges, 
distributional limit theorems for the number of events and occupation times are shown analytically. These are known as the Mittag-Leffler distribution and the 
generalized arcsine law in probability theory.
\end{abstract}

\maketitle


\section{Introduction}

Fluctuations are essential and often contain useful and important information in nature. In stochastic thermodynamics, fluctuations of the entropy production 
follow the fluctuation theorem \cite{Evans1993, Gallavotti1995}, which gives the Jarzynski equality and the  second law of the thermodynamics \cite{Jarzynski1997, Seifert_2012}. 
In diffusion processes, fluctuations of the displacement of a particle, i.e., the variance of the displacement,
 increases linearly with time. The degree of diffusivity is characterized by the proportional constant, i.e., 
the diffusion coefficient. Furthermore, in random search processes, some fluctuations of diffusive modes optimize a searching time \cite{Benichou2010, Benichou2011}. 
Moreover, in a biological system, fluctuations of amino acids regulate water transportation in aquaporin 1 \cite{Yamamoto2014b}.
Therefore, fluctuations play an important role in natural phenomena.  

There are some universal fluctuations in stochastic processes. The central limit theorem provides one of the most universal fluctuations, i.e., the normal distribution 
\cite{Feller1971}, which states that the normalized sum of independent and identically distributed (IID) random variables converges in distribution to the normal distribution. 
The central limit theorem is valid for any IID random variables if there exists a finite second moment. When the second moment diverges, the generalized central limit 
theorem holds, which states that the normalized sum of IID random variables converges in distribution to the stable distribution \cite{Feller1971}. 
Another classical probability theory is the arcsine law, which states that
 the occupation time distribution in the positive side in a random walk or the Brownian motion follows the arcsine distribution \cite{Feller1971}.
Its generalization is known as the generalized arcsine law \cite{Dynkin1961}. The occupation time distribution depends on the domain. When the domain is finite, 
the occupation time distribution in the Brownian motion 
follows the half Gaussian, and its generalization to general Markov processes provides the Mittag-Leffler distribution \cite{Darling1957}. These universal fluctuations 
play a fundamental role in infinite ergodic theory \cite{Aaronson1981, Aaronson1997, Thaler1998, Thaler2002, Akimoto2008, Akimoto2015}.

In biological and soft mater systems, diffusivity often changes randomly. This fluctuating diffusivity
 provides non-Gaussian fluctuations in the displacement distribution, anomalous diffusion, and ergodicity breaking 
\cite{Doi-Edwards-book, Yamamoto-Onuki-1998, Richert-2002, wang2012brownian, Manzo2015, Yamamoto2021}. 
Brownian motion with fluctuating diffusivity is a simple stochastic model of such systems \cite{Chubynsky2014, Massignan2014, Uneyama2015, Chechkin2017}.
These stochastic models show a non-Gaussian distribution in the displacement, anomalous diffusion, and non-ergodic behaviors in the time-averaged diffusivity 
\cite{Chubynsky2014, Massignan2014, Uneyama2015, Chechkin2017, Miyaguchi2016, AkimotoYamamoto2016a}.
As a simple stochastic model of the Brownian motion with fluctuating diffusivity, 
Brownian motion with dichotomously fluctuating diffusivity is investigated \cite{Miyaguchi2016, Akimoto-Yamamoto2016}, where 
low and high diffusivities change alternatively. In particular, such a fluctuating diffusivity is modeled by a dichotomous process.

Renewal process is a point process where the duration times of a state are IID random variables. Occupation time statistics and the statistics of the number of renewals 
are studied in physics and mathematics literatures because of its wide applications \cite{God2001, Margolin2005, Margolin2006, lefevere2011large, Akimoto2012, Zaburdaev2015, horii2022large, Cox, Feller1971}. 
An alternating renewal processes is a slight modification of 
the renewal process, where duration times are IID random variables but the duration-time distributions alternates \cite{Cox,mitov2014renewal}. 
In this paper, we aim to obtain occupation time statistics, the moments of the number of changes of states, and the correlation function 
in alternating renewal processes. These theoretical results play an important role in fundamental theory as well as numerous applications to physical systems 
 such as the Langevin equation with fluctuating diffusivity  \cite{Miyaguchi2016, Akimoto-Yamamoto2016}, 
 the mean magnetization in spin systems \cite{God2001}, the fluorescence of quantum dots \cite{Brok2003}, interface fluctuations in 
liquid crystals \cite{takeuchi2016characteristic}, 
$\alpha$-percentile options in stock prices \cite{miura1992note, akahori1995some}, and leads in sports games \cite{Clauset2015}.
In particular, occupation time statistics of one of two diffusive states are important in obtaining the mean square displacement 
in the Brownian motion with dichotomously fluctuating diffusivity. Moreover, the statistics of the number of changes of states play an important role 
in the continuous-time random walk and its generalization \cite{He2008, Miyaguchi2011, Miyaguchi2013}.

This paper is organized as follows. In section II, we describe an alternating renewal process and observables that we are interested in.
In section III, we derive the backward and the forward recurrence time distributions. In section IV, we derive the moments of the number of renewals. 
In section V, we show the occupation time statistics. In section VI, we derive the correlation function of states. Section VII is devoted to the conclusion. 

\section{model and observables}

Renewal process is a point process where an inter-event time between successive renewals is an IID random variable.
If there are two types of renewal processes, the inter-event-time distributions for the renewal processes take different forms. In particular, if the inter-event time distribution alternates between 
the two distributions, this process is called an alternating renewal process. 
We consider a dichotomous random signal $\sigma(t)$ described by an alternating renewal processes. The random signal $\sigma(t)$ alternates 
between  $+$ and $-$ states, i.e., $\sigma(t)=+1$ or $-1$. 
Duration times for $+$ and $-$ states are 
random variables following different probability density functions (PDFs), $\rho_+(x)$ and $\rho_-(x)$ 
for $+$ and $-$ states, respectively. 
Here, we consider four cases for the PDFs of duration times. These cases are summarized by the Laplace transform of the PDFs:
\begin{enumerate}
\item $\alpha_+ \leq \alpha_- <1$: $\hat{\rho}_\pm(s) = 1 - a_\pm s^{\alpha_\pm} + o(s^{\alpha_\pm})$,
\item $1<\alpha_+ < \alpha_-<2$: $\hat{\rho}_\pm(s) = 1 - \mu_\pm s + a_\pm s^{\alpha_\pm} + o(s^{\alpha_\pm})$,
\item $2<\alpha_\pm$: $\hat{\rho}_\pm(s) = 1 - \mu_\pm s+ \frac{1}{2}(\sigma^2_\pm + \mu_\pm^2)s^2 + o(s^2)$,
\item $\alpha_+<1$ and $1<\alpha_-$: $\hat{\rho}_+(s) = 1 - a_+ s^{\alpha_+} + o(s^{\alpha_+})$ and $\hat{\rho}_-(s) = 1 - \mu_- s + o(s)$,
\end{enumerate}
where $\hat{\rho}_\pm(s)$ is the Laplace transform of $\rho_\pm(t)$, $\mu_\pm$ is the mean duration time, and $\sigma_\pm^2$ is the variance. 
In what follows, we use $\alpha\equiv \min(\alpha_+,\alpha_-)$.
In Case 1, both of the mean duration times diverge.  
In Case 2, both of the mean duration times are finite and both second moments of duration times diverge. In Case 3, 
both second moments of duration times are finite. In Cases 1,2, and 3, the asymptotic behavior 
of the PDFs follow power-law distributions: $\rho_\pm(x) \propto x^{-1-\alpha_\pm}$ for $x\to \infty$. 
In Case 4, the asymptotic behavior of the PDF of duration times for $+$ state follows power-law distributions: $\rho_+(x) \propto x^{-1-\alpha_+}$ for $x\to \infty$. 
Thus, the mean duration time for $+$ state diverges. Moreover, the asymptotic behavior of the PDF of duration times for $-$ state follows a  power-law distribution: 
$\rho_-(x) \propto x^{-1-\alpha_-}$ for $x\to \infty$.


\begin{figure}
\includegraphics[width=.9\linewidth, angle=0]{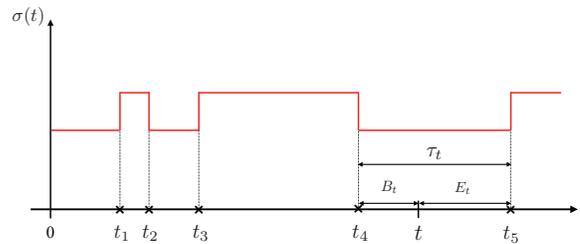}
\caption{Renewal time $t_n$, the forward recurrence time $E_t$, the backward recurrence time $B_t$, and the time interval straddling $t$, i.e., $\tau_t$.}
\label{observables}
\end{figure}

We consider the following observables, which are illustrated in Figure~\ref{observables}.
 $t_n$ is the time at $n$th renewal, i.e, $t_n=\tau_1 + \tau_2 + \cdots +\tau_n$, where $\tau_k$ is the $k$th duration time. 
 $N_t$ is the number of renewals from 0 to $t$. 
 $E_t$ is the forward recurrence time defined by $E_t \equiv  t_{N_{t+1}} - t$.
 $B_t$ is the backward recurrence time defined by $B_t \equiv  t -t_{N_t}$.
 $\tau_t$ is the time interval straddling $t$. 
 $T_t^{\pm}$ is the occupation time for the $\pm$ state, which is represented by 
\begin{equation}
T_t^+ = \frac{1}{2} \int_0^t (1 + \sigma(t'))dt'.
\end{equation}
When $\sigma (0)=+1$, $T_t^+ = \tau_1 + \tau_3 + \cdots +\tau_{2k+1}$  
or $T_t^+ = \tau_1 + \tau_3 + \cdots +\tau_{2k-1} + B_t$ if $N_t = 2k+1$ or $N_t = 2k$, respectively. 
Moreover, when $\sigma (0)=-1$, $T_t^+ = \tau_2 + \tau_4 + \cdots +\tau_{2k} + B_t$ 
or $T_t^+ = \tau_2 + \tau_4 + \cdots +\tau_{2k}$ if $N_t = 2k+1$ or $N_t = 2k$, respectively. 
In what follows, we assume $\sigma(0)=+1$ except where specifically noted. 
It follows that the occupation time for the $-$ state with $\sigma(0)=\pm1$, denoted by $T_t^-(\sigma_0=\pm1)$, can be represented by 
$T_t^-(\sigma_0=+1) = T_t^+(\sigma_0=-1)$  and $T_t^-(\sigma_0=-1) = T_t^+(\sigma_0=+1)$.

\section{Backward and Forward recurrence time distributions}

\subsection{forward recurrence time distribution}

We derive the forward recurrence time distribution. 
It is intuitively conjectured that the probability finding $+$ state, $P_+$, and $-$ state, $P_-$,  for $t\to \infty$ 
can be represented by the means, i.e., $P_+=\mu_+/(\mu_+ + \mu_-)$ and $P_-=\mu_-/(\mu_+ + \mu_-)$, respectively, if the means exist. 
Furthermore, the probabilities do not depend on the initial condition. This is rigorously proved in Ref.~\cite{Cox}.
We assume that the PDF of the duration time for the first renewal is the 
same as $\rho_+(\tau)$ or  $\rho_-(\tau)$. This process is called an {\it ordinary alternating renewal process} \cite{Cox}.
The joint PDF of $E_t$ and $N_t=2n$ ($n=0,1, \cdots$) for fixed $t$ can be represented by 
\begin{eqnarray}
f_{E,2n}(E_t;t) 
= \langle \delta (E_t - t_{2n+1} + t) I(t_{2n} < t < t_{2n+1})\rangle,
\end{eqnarray}
where $\langle \cdot \rangle$ is the expectation value. 
The double Laplace transform of $f_{2n}(E_t;t)$ with respect to $E_t$ and $t$ is defined by
\begin{widetext}
\begin{equation}
\hat{f}_{E,2n}(u;s) \equiv \left\langle \int_0^\infty dE_t e^{-uE_t} \int_0^\infty dt e^{-st}  \delta (E_t - t_{2n+1} + t) I(t_{2n} < t < t_{2n+1}) \right\rangle.  
\end{equation}
\end{widetext}
A simple calculation yields 
\begin{equation}
\hat{f}_{E,2n}(u;s) = \{ \hat{\rho}_+(s) \hat{\rho}_-(s) \}^n \frac{\hat{\rho}_+(u) - \hat{\rho}_+(s)}{s-u}.
\end{equation}
In the same way, we have the double Laplace transform of $f_{E,2n+1}(E_t;t)$ ($n=0,1, \cdots$):
\begin{equation}
\hat{f}_{2n+1}(u;s) = \{ \hat{\rho}_+(s) \hat{\rho}_-(s) \}^n \hat{\rho}_+(s) \frac{\hat{\rho}_-(u) - \hat{\rho}_-(s)}{s-u}.
\end{equation}
The double Laplace transforms of the PDFs of $E_t$ with final states being $+$ and $-$,  denoted by $\hat{f}_{E,+}(u;s)$ and $\hat{f}_{E,-}(u;s)$, are given by
$\hat{f}_{E,+}(u;s) = \sum_{n=0}^\infty \hat{f}_{E,2n}(u;s)$ and $\hat{f}_{E,-}(u;s)=\sum_{n=0}^\infty \hat{f}_{E,2n+1}(u;s)$, respectively. Note that we assumed 
$\sigma(0)=+1$. 
We obtain
\begin{equation}
\hat{f}_{E,+}(u;s)=\frac{\hat{\rho}_+(u) - \hat{\rho}_+(s)}{(s-u)\{1 - \hat{\rho}_+(s) \hat{\rho}_-(s)\}}
\label{Laplace P+}
\end{equation}
and
\begin{equation}
\hat{f}_{E,-}(u;s)=\frac{\hat{\rho}_+(s) \{\hat{\rho}_-(u) - \hat{\rho}_-(s)\}}{(s-u)\{1 - \hat{\rho}_+(s) \hat{\rho}_-(s)\}}.
\label{Laplace P-}
\end{equation}
The double Laplace transform of the forward recurrence time distribution becomes 
\begin{eqnarray}
\hat{f}_E(u;s) 
&=&  \sum_{n=0}^\infty \hat{f}_{E,2n}(u;s) + \sum_{n=0}^\infty \hat{f}_{E,2n+1}(u;s) \\
&=& \frac{ \hat{\rho}_+(u) - \hat{\rho}_+(s) + \hat{\rho}_+(s) \{ \hat{\rho}_-(u) - \hat{\rho}_-(s)\}}{(s-u) \{ 1 - \hat{\rho}_+(s) \hat{\rho}_-(s)\}}.
\end{eqnarray}
We note that the result with $\rho_+(x)=\rho_-(x)$ is consistent with Ref.~\cite{God2001}.  
In the same way,  we have the double Laplace transform of the forward recurrence time distribution in the case  of $\sigma(0)=-1$:
\begin{equation}
\hat{f}_E(u;s) 
= \frac{ \hat{\rho}_-(u) - \hat{\rho}_-(s) + \hat{\rho}_-(s) \{ \hat{\rho}_+(u) - \hat{\rho}_+(s)\}}{(s-u) \{ 1 - \hat{\rho}_+(s) \hat{\rho}_-(s)\}}.
\end{equation}
For $s\to 0$, the double Laplace transforms  $\hat{f}_E(u;s)$ for both initial conditions coincide and are given by
\begin{equation}
\hat{f}_E(u;s) \sim \frac{ \hat{\rho}_-(u) - \hat{\rho}_-(s) +  \hat{\rho}_+(u) - \hat{\rho}_+(s) }{(s-u) \{ 1 - \hat{\rho}_+(s) \hat{\rho}_-(s)\}}.
\label{double Laplace Et} 
\end{equation}
Therefore,  the PDF of the forward recurrence time does not depend on the initial state in the long-time limit ($t\to\infty$).

\subsubsection{Probabilities finding $+$ and $-$ states}
By Eqs.~(\ref{Laplace P+}) and (\ref{Laplace P-}), we have the probabilities finding $+$ and $-$ states, $P_+$ and $P_-$, for $t\to \infty$. 
Taking limits of  $s\to 0$ and $u= 0$  in Eqs.~(\ref{Laplace P+}) and (\ref{Laplace P-}) yields the probabilities.
For Cases 2 and 3 ($\alpha >1$), the probabilities become 
\begin{equation}
P_+ =\frac{\mu_+}{\mu_+ + \mu_-}\quad {\rm and}\quad P_-=\frac{\mu_-}{\mu_+ + \mu_-}.
\label{prob_+-_a>1}
\end{equation}
These results are consistent with the intuitive understanding. 
On the other hand, for $\alpha_+ < \alpha_- < 1$ (Case 1), the probability finding $-$ state decays to zero in the long-time limit. 
For $\alpha_+ < \alpha_- < 1$ (Case 1), the asymptotic behavior of the probability $P_-(t)$ of finding $-$ state at time $t$ is given by  
\begin{equation}
P_-(t) \sim \frac{a_-}{a_+ \Gamma(1+\alpha_+-\alpha_-)} \frac{1}{t^{\alpha_--\alpha_+}}.
\end{equation}
Moreover, for  $\alpha_+=\alpha_-<1$ (Case 1), the probabilities converge to a finite value:
\begin{equation}
P_+ =\frac{a_+}{a_+ + a_-}\quad {\rm and}\quad P_-=\frac{a_-}{a_+ + a_-}.
\end{equation}
For $\alpha_+ < 1$ and $1<\alpha_- $ (Case 4), the probability of finding a $-$ state also decays to zero in the long-time limit and the asymptotic behavior is given by  
\begin{equation}
P_-(t) \sim \frac{\mu_-}{a_+ \Gamma(\alpha)} \frac{1}{t^{1-\alpha}}.
\end{equation}

\subsubsection{Asymptotic behavior of the forward recurrence time distribution}
For Cases 1 and 4, the double Laplace transform of the forward recurrence time distribution for $s\ll 1$ and $u\ll 1$ with $s/u=O(1)$ becomes
\begin{eqnarray}
\hat{f}_E(u;s) &\cong& \frac{u^\alpha - s^\alpha}{(u-s)s^\alpha},
\end{eqnarray}
which is the exactly same as in the case of $\rho_+(x)=\rho_-(x)$ \cite{God2001}. Using an inversion method of Ref.~\cite{God2001}, we have the PDF of $E_t/t$ for Cases 1 and 4: 
\begin{equation}
\lim_{t\to\infty} f_{E_t/t}(x) = \frac{\sin \pi \alpha}{\pi} \frac{1}{x^\alpha (1+x)}.
\label{fwd recurrence time dist alpha<1}
\end{equation}
For Cases 2 and 3, i.e., $\hat{\rho}_\pm (s) = 1 - \mu_\pm s  + o(s)$, in the long time limit ($t\to\infty$), the Laplace transform of the PDF of $E_t$ reads 
\begin{eqnarray}
\lim_{t\to\infty} \hat{f}_E(u;t) &=& \lim_{s\to0} s\hat{f}_E(u;s) \nonumber\\
&=&\frac{2-\hat{\rho}_+(u) - \hat{\rho}_-(u)}{(\mu_++\mu_-)u}\\
&=& P_+ \hat{f}_{E,+}(u) +  P_{-} \hat{f}_{E,-}(u),
\end{eqnarray}
where 
\begin{equation}
\hat{f}_{E,+}(u) = \frac{1-\hat{\rho}_+(u)}{\mu_+ u} ~ {\rm and}~ \hat{f}_{E,-}(u) = \frac{1-\hat{\rho}_-(u)}{\mu_- u}.
\end{equation}
By the inverse Laplace transformation, we have the PDF of $E_t$  in the long time limit ($t\to\infty$) for cases 2 and 3:
\begin{equation}
f(x) = \frac{1}{\mu_++\mu_-} \left[ \int_x^\infty \rho_+(\tau)d\tau + \int_x^\infty \rho_- (\tau)d\tau \right].
\label{forward_PDF}
\end{equation}
When the PDF of the duration time for the first renewal is given by Eq.~(\ref{forward_PDF}). This process is called an {\it equilibrium alternating renewal process} \cite{Cox}.
In the equilibrium alternating renewal process, the probability of finding $\sigma(0)=1$ is given by $P_+$.

The mean of $E_t$ in the limit $t\to\infty$ diverges in Case 2 whereas the mean duration time is finite. Here, we calculate a long-time behavior of 
the mean of $E_t$. The Laplace transform of $\langle E_t \rangle$ with respect to $t$ is given by 
\begin{equation}
\mathcal{L}(\langle E_t \rangle ) \equiv \int_0^\infty dt e^{-st} \langle E_t \rangle = \left. - \frac{\partial \hat{f}_E(u;s)}{\partial u} \right|_{u=0} ,
\end{equation}
which becomes 
\begin{equation}
\mathcal{L}(\langle E_t \rangle ) 
=\frac{2 - \rho_+(s) - \rho_-(s) - (\mu_+ + \mu_-)s}{s^2 \{ 1 - \rho_+(s) \rho_-(s)\}}. 
\end{equation}
In the case of $1<\alpha_+=\alpha_-<2$ (Case 2), we have 
\begin{equation}
\int_0^\infty dt e^{-st} \langle E_t \rangle \sim \frac{a_++a_-}{\mu_+ + \mu_-} \frac{1}{s^{3-\alpha}} \quad (s\to 0).
\end{equation}
By the inverse Laplace transform, the asymptotic behavior  of $\langle E_t \rangle$ for $1<\alpha_+=\alpha_-<2$ (Case 2) becomes 
\begin{equation}
\langle E_t \rangle \sim \frac{a_++a_-}{(\mu_+ + \mu_-)\Gamma(3-\alpha)} t^{2-\alpha} \quad (t\to \infty).
\end{equation}

\subsection{backward recurrence time distribution}
Here we calculate the backward recurrence time distribution, which is almost the same as the calculation of the forward recurrence time distribution.
The joint PDF of $B_t$ and $N_t=2n$ is given by
\begin{equation}
f_{B,2n}(B_t;t) = \langle \delta (B_t - t + t_{2n}) I(t_{2n} < t < t_{2n+1})\rangle.
\end{equation}
The double Laplace transforms of $f_{B,2n}(B_t;t)$ and $f_{B,2n+1}(B_t;t)$ with respect to $B_t$ and $t$ are given by
\begin{equation}
\hat{f}_{B,2n}(u;s) = \{ \hat{\rho}_+(s) \hat{\rho}_-(s) \}^n \frac{1 - \hat{\rho}_+(s+u)}{s+u},
\end{equation}
and
\begin{equation}
\hat{f}_{B,2n+1}(u;s) = \{ \hat{\rho}_+(s) \hat{\rho}_-(s) \}^n \rho_+(s) \frac{1 - \hat{\rho}_-(s+u)}{s+u}.
\end{equation}
It follows that the double  Laplace transform of the PDF of $B_t$ is given by 
\begin{eqnarray}
\hat{f}_{B}(u;s) &=& \sum_{n=0}^\infty \hat{f}_{B,2n}(u;s) + \sum_{n=0}^\infty \hat{f}_{B,2n+1}(u;s)\\
&=& \frac{1 - \hat{\rho}_+(s+u) + \hat{\rho}_+(s) \{ 1 - \hat{\rho}_-(s+u)\}}{(s+u)\{1 -\hat{\rho}_+(s) \hat{\rho}_-(s)\}}.
\end{eqnarray}
For $\alpha>1$ (Cases 2 and 3), in the long time limit ($t\to\infty$), the Laplace transform of the PDF of $B_t$ reads 
\begin{eqnarray}
\lim_{t\to\infty} \hat{f}_B(u;t) &=& \lim_{s\to0} s\hat{f}_B(u;s)=\frac{2-\hat{\rho}_+(u) - \hat{\rho}_-(u)}{(\mu_++\mu_-)u}\\
&=& P_+ \hat{f}_{E,+}(u) + P_{-} \hat{f}_{E,-}(u).
\end{eqnarray}
Therefore, the backward recurrence time distribution is the same as the forward recurrence time distribution when $\alpha>1$.
On the other hand, for $\alpha <1$ (Cases 1 and 4), the Laplace transform for $s\ll 1$ and $u\ll 1$ with $s/u=O(1)$ becomes 
\begin{equation}
\hat{f}_B(u;s) \cong s^{-\alpha} (s+u)^{\alpha-1}, 
\end{equation}
which is the exactly same as in the case of $\rho_+(x)=\rho_-(x)$. An inversion method in Ref.~\cite{God2001} yields
\begin{equation}
\lim_{t\to\infty} f_{B_t/t}(x) = \frac{\sin \pi \alpha}{\pi} x^{-\alpha}(1-x)^{\alpha -1}.
\end{equation}

\subsection{distribution of the time interval straddling $t$}
Here we calculate the distribution of the time interval straddling $t$, i.e., $\tau_t$ \cite{Barkai2014}. Counter-intuitively, this distribution is not the same as 
$P_+ \rho_+(x) + P_- \rho_-(x)$. 
The joint PDF of $\tau_t$ and $N_t=2n$ is given by
\begin{equation}
f_{2n}(\tau_t;t) = \langle \delta (\tau_t - t_{2n+1} + t_{2n}) I(t_{2n} < t < t_{2n+1})\rangle.
\end{equation}
The double Laplace transforms of $f_{2n}(\tau_t;t)$ and $f_{2n+1}(\tau_t;t)$ with respect to $\tau_t$ and $t$ are given by
\begin{equation}
\hat{f}_{2n}(u;s) = \{ \hat{\rho}_+(s) \hat{\rho}_-(s) \}^n \frac{\hat{\rho}_+(u) - \hat{\rho}_+(s+u)}{s},
\end{equation}
and
\begin{equation}
\hat{f}_{2n+1}(u;s) = \{ \hat{\rho}_+(s) \hat{\rho}_-(s) \}^n \hat{\rho}_+(s) \frac{\hat{\rho}_-(u) - \hat{\rho}_-(s+u)}{s}.
\end{equation}
It follows that the double  Laplace transform of the PDF of $\tau_t$ is given by 
\begin{equation}
\hat{f}(u;s) 
= \frac{ \hat{\rho}_+(u) - \hat{\rho}_+(s+u) + \hat{\rho}_+(s) \{ \hat{\rho}_-(u) - \hat{\rho}_-(s+u)\}}{s\{1 -\hat{\rho}_+(s) \hat{\rho}_-(s)\}}.
\end{equation}
For $\alpha>1$ (Cases 2 and 3), in the long time limit ($t\to\infty$), the Laplace transform of the PDF of $\tau_t$ reads 
\begin{eqnarray}
\lim_{t\to\infty} \hat{f}(u;t) &=& \lim_{s\to0} s\hat{f}(u;s) \\ 
&=& -\frac{ \hat{\rho}_+'(u) +  \hat{\rho}_-'(u)}{\mu_+ + \mu_-}.
\end{eqnarray}
Therefore, the distribution of the time interval straddling $t$ is not the same as $P_+ \rho_+(x) + P_- \rho_-(x)$. 
For $\alpha <1$ (Cases 1 and 4), the Laplace transform for $s\ll 1$ and $u\ll 1$ with $s/u=O(1)$ becomes 
\begin{equation}
\hat{f} (u;s) \cong \frac{(s+u)^{\alpha} - s^\alpha}{s^{1+\alpha}}, 
\end{equation}
which is the same as in the case of $\rho_+(x)=\rho_-(x)$. An inversion method in Ref.~\cite{God2001} yields
\begin{eqnarray}
&&\lim_{t\to\infty} f_{\tau_t/t}(x) \nonumber\\
&=&\left\{
\begin{array}{ll}
 \dfrac{\sin \pi \alpha}{\pi x^{1+\alpha}} [1 - (1-x)^{\alpha -1} ] \quad &(x<1)\\
 \\
 \dfrac{\sin \pi \alpha}{\pi x^{1+\alpha}} [1 - 2(x-1)^{\alpha -1} \cos \pi \alpha] \quad &(x<1).
 \end{array}
 \right.
\end{eqnarray}

\section{The moments of the number of renewals}
Here, we consider the moments of the number of renewals in the time interval $(0,t)$, i.e., $N_t$. 

\subsection{First moment}
The renewal function, which is the mean of $N_t$,
can be obtained as 
\begin{eqnarray}
H(t)\equiv \langle N_t \rangle &=& \sum_{r=0}^\infty r \Pr (N_t=r) \nonumber\\
&=& \sum_{r=0}^\infty r \{\Pr (t_r<t) - \Pr(t_{r+1}<t)\} \nonumber\\
&=&\sum_{r=1}^\infty \Pr(t_r<t).
\end{eqnarray}
Taking the Laplace transform with respect to $t$ yields 
\begin{equation}
\hat{H}(s) = \frac{1}{s}\sum_{r=1}^\infty \hat{\rho}_1(s) \cdots \hat{\rho}_r(s),
\end{equation}
where $\hat{\rho}_r (s)$ is the Laplace transform for the $r$th duration-time PDF. In what follows, we consider three alternating renewal 
processes: equilibrium, ordinary, and aging alternating renewal processes. In the ordinary alternating renewal process, $\rho_1(x)$ is the same as $\rho_+(x)$, where we assume 
that the initial state is $\sigma(0)=1$. In the equilibrium alternating renewal process, $\rho_1(x)$ is given by Eq.~(\ref{forward_PDF}) and the probabilities finding $\pm$ at $t=0$ 
are given by Eq.~(\ref{prob_+-_a>1}). We note that the equilibrium alternating renewal process exists only if $\alpha>1$. For $\alpha \leq 1$, there is no equilibrium distribution 
in the forward recurrence time. In this case, the statistical properties in the time interval $[t_a, t_a + t]$ explicitly depend on the aging time $t_a$ 
\cite{bouchaud1995aging, God2001,Schulz2013, Schulz2014, Akimoto2013}. This process is called the {\it aging alternating renewal process}. 

\subsubsection{ordinary alternating renewal process}
In the ordinary alternating renewal process, we have
\begin{equation}
\hat{H}(s) = \frac{\hat{\rho}_+(s)[1+\hat{\rho}_-(s)]}{s[1-\hat{\rho}_+(s) \hat{\rho}_-(s)]}.
\end{equation}
The leading order is given by 
\begin{widetext}
\begin{equation}
\hat{H}(s) = \left\{
\begin{array}{ll}
\dfrac{2}{(\mu_+ + \mu_-)s^2} + \dfrac{\sigma_+^2 + \sigma_-^2 - \mu_-^2 - \mu_+\mu_-}{(\mu_+ + \mu_-)^2s} + o(s^{-1}) \quad (2< \alpha ,~Case~3),\\
\\
\dfrac{2}{(\mu_+ + \mu_-)s^2} + \dfrac{a_+}{(\mu_+ + \mu_-)^2s^{3-\alpha}} + o(s^{-3-\alpha}) \quad (1<\alpha < 2,~Case~2),\\
\\
\dfrac{2}{(a_++a_-)s^{\alpha+1}} + o(s^{-1-\alpha}) \quad (\alpha_+=\alpha_-<1,~Case~1),\\
\\
\dfrac{2}{a_+s^{\alpha+1}} + o(s^{-1-\alpha}) \quad (\alpha_+<1,~Cases~1~and~4).\\
\end{array}
\right.
\end{equation}
In the long-time limit, the renewal function becomes 
\begin{equation}
H(t) = \left\{
\begin{array}{ll}
\dfrac{2t}{\mu_+ + \mu_-} + \dfrac{\sigma_+^2 + \sigma_-^2 - \mu_-^2 - \mu_+\mu_-}{(\mu_+ + \mu_-)^2} + o(1)\quad (2 < \alpha ,~Case~3),\\
\\
\dfrac{2t}{\mu_+ + \mu_-} + \dfrac{a_+}{(\mu_+ + \mu_-)^2 \Gamma(3-\alpha)}t^{2-\alpha} + o(t^{2-\alpha}) \quad (1<\alpha < 2,~Case~2),\\
\\
\dfrac{2}{(a_++a_-)\Gamma(\alpha+1)} t^\alpha + o(t^\alpha) \quad (\alpha_+=\alpha_-<1,~Case~1),\\
\\
\dfrac{2}{a_+\Gamma(\alpha+1)} t^{\alpha} + o(t^\alpha) \quad (\alpha_+<\alpha_-<1,~Cases~1~and~4).
\end{array}
\right.
\label{renewal function}
\end{equation}
\end{widetext}
When the mean duration time diverges, the renewal function increases sublinearly in the asymptotic behavior. 
This is a mechanism of subdiffusion in the continuous-time random walk \cite{metzler00, klafter2011first}.

\subsubsection{equilibrium alternating renewal process}
In equilibrium renewal process ($\alpha > 1$), the PDF of the first renewal time 
PDF is given by Eq.~(\ref{forward_PDF}). More precisely, if the initial state is $+$, the PDF of the first renewal time is given by
\begin{equation}
f_+(\tau) = \frac{1}{\mu_+} \int_\tau^\infty \rho_+(x)dx,
\end{equation} 
and 
\begin{equation}
f_-(\tau) = \frac{1}{\mu_-} \int_\tau^\infty \rho_+(x)dx,
\end{equation} 
otherwise. In the equilibrium process, the Laplace transform of the renewal function is exactly obtained as
\begin{equation}
\hat{H}(s)  = \frac{2}{(\mu_++\mu_-)s^2}.
\end{equation}
By the inverse Laplace transform, the renewal function for Cases 2 and 3 becomes 
\begin{equation}
H(t)=\frac{2}{ \mu_++\mu_-} t.
\end{equation}
Therefore, the renewal function can be represented by the mean duration time $\mu=(\mu_+ + \mu_-)/2$, i.e., $H(t)=t/\mu$, which is consistent with the intuition.

\subsubsection{aging alternating renewal process}

For $\alpha<1$ (Cases 1 and 4), there is no equilibrium distribution in the forward recurrence time. As a result, the forward recurrence time distribution explicitly depends on the 
elapsed time $t_a$ (aging time) of the system, where the ordinary alternating renewal process is assumed at time $t=0$. 
By Eq.~(\ref{fwd recurrence time dist alpha<1}), the asymptotic behavior of the forward recurrence time distribution 
for $t\gg 1$ becomes
\begin{equation}
 f_{E_t}(x) \sim \frac{\sin \pi \alpha}{\pi } \frac{t^\alpha}{x^\alpha (t+x)}.
\end{equation}
The asymptotic behavior of the double Laplace transform of the renewal function $H(t; t_a)$, which is the mean number of renewals in $[t_a, t_a + t]$, 
with respect to $t$ and the aging time $t_a$ ($t_a \leftrightarrow u$ and $t \leftrightarrow s$) is approximately given by
\begin{equation}
\hat{H}(s; u)  \cong \frac{ 2\hat{f}_{E}(s;u)   }{s [1-\hat{\rho}_+(s) \hat{\rho}_-(s)]}.
\end{equation}
For $\alpha_+<\alpha_-<1$ (Cases 1 and 4), the leading order becomes 
\begin{equation}
\hat{H}(s; u) \sim \left\{
\begin{array}{ll}
\dfrac{2}{a_+ s^{1+\alpha} u}\quad (s\ll u),\\
\\
\dfrac{2}{a_+s^{2} u^{\alpha}} \quad (s \gg u).\\
\end{array}
\right.
\end{equation}
By the inverse Laplace transform, the asymptotic behavior of $H(t; t_a)$ for $\alpha_+<\alpha_-<1$ (Cases 1 and 4) becomes 
\begin{equation}
H(t; t_a) \sim \left\{
\begin{array}{ll}
\dfrac{2t^{\alpha}}{a_+ \Gamma (1+\alpha)}\quad (t \gg t_a),\\
\\
\dfrac{2 t t_a^{\alpha -1}}{a_+  \Gamma(\alpha)} \quad (t \ll t_a).\\
\end{array}
\right.
\end{equation}

\subsection{second moment}
The second moment of $N_t$ is also obtained as 
\begin{eqnarray}
H_2 (t) \equiv \langle N_t ^2\rangle &=& \sum_{r=0}^\infty r^2 \Pr (N_t=r)\\
&=&\sum_{r=1}^\infty (2r-1) \Pr(t_r<t).
\end{eqnarray}
Taking the Laplace transform with respect to $t$ yields 
\begin{equation}
\hat{H}_2(s) = \frac{1}{s}\sum_{r=1}^\infty (2r-1) \hat{\rho}_1(s) \cdots \hat{\rho}_r(s).
\end{equation}

\subsubsection{ordinary alternating renewal process}
In the ordinary renewal process, we have
\begin{widetext}
\begin{equation}
\hat{H}_2(s) =  \frac{\hat{\rho}_+(s)[1 +  3 \hat{\rho}_-(s) + 3 \hat{\rho}_-(s) \hat{\rho}_+(s) + \hat{\rho}_-(s)^2 \hat{\rho}_+(s) ]}{ s [1- \hat{\rho}_+(s) \hat{\rho}_-(s)]^2}.
\end{equation}
The asymptotic behaviors are given by 
\begin{equation}
\hat{H}_2(s) = \left\{
\begin{array}{ll}
\dfrac{8}{(\mu_+ + \mu_-)^2 s^3} + \dfrac{8(\sigma_+^2 + \sigma_-^2) -4 \mu_+ (\mu_+ +\mu_-)  }{(\mu_+ + \mu_-)^3 s^2}  + o(s^{-2}) \quad (2< \alpha ,~Case~3),\\
\\
\dfrac{8}{(\mu_+ + \mu_-)^2 s^3} + \dfrac{16a_+}{(\mu_+ + \mu_-)^3 s^{4-\alpha}}+ o(s^{\alpha-4}) \quad (1<\alpha < 2,~Case~2),\\
\\
\dfrac{8}{a_+^2 s^{2\alpha+1}}  + o(s^{-1-\alpha}) \quad (\alpha_+<\alpha_-<1,~Cases~1~and~4).\\
\end{array}
\right.
\end{equation}
In the long-time limit, the second moment of $N_t$ becomes 
\begin{equation}
H_2(t) = \left\{
\begin{array}{ll}
\dfrac{4t^2}{(\mu_+ + \mu_-)^2} + \dfrac{8(\sigma_+^2 + \sigma_-^2) -4 \mu_+ (\mu_+ +\mu_-)  }{(\mu_+ + \mu_-)^3 }t +o(t) \quad (2< \alpha ,~Case~3),\\
\\
\dfrac{4t^2}{(\mu_+ + \mu_-)^2} + \dfrac{16a_+}{(\mu_+ + \mu_-)^3 \Gamma(4-\alpha)} t^{3-\alpha} +o(t^{3-\alpha}) \quad (1< \alpha <2,~Case~2),\\
\\
\dfrac{8}{a_+^2 \Gamma(2\alpha+1)} t^{2\alpha} \quad (\alpha_+<\alpha_-<1,~Cases~1~and~4).
\end{array}
\right.
\end{equation}
The variance of $N_t$ is given by
\begin{equation}
{\rm Var}(N_t)  \sim \left\{
\begin{array}{ll}
\dfrac{4(\sigma_+^2 + \sigma_-^2) -4  (\mu_+^2 - \mu_-^2)  }{(\mu_+ + \mu_-)^3 }t  \quad (2<\alpha ,~Case~3),\\
\\
\dfrac{4a_+ (1+\alpha) }{(\mu_+ + \mu_-)^3 \Gamma(4-\alpha)} t^{3-\alpha} \quad (1<\alpha<2 ,~Case~2),\\
\\
\dfrac{4(2 \Gamma(\alpha+1)^2 - \Gamma(2\alpha +1))} {a_+^2 \Gamma(2\alpha +1)\Gamma(\alpha+1)^2} t^{2\alpha} \quad (\alpha_+<\alpha_-<1,~Cases~1~and~4).
\end{array}
\right.
\label{variance Nt ord}
\end{equation}
For $1<\alpha<2$, the variance of $N_t$ increases as $t^{3-\alpha}$, where $3-\alpha >1$. Therefore, the variance grows faster than 
that for $\alpha>2$. This is a mechanism of the field-induced superdiffusion \cite{Gradenigo2016,Hou2018,Akimoto2018b}.

\subsubsection{equilibrium alternating renewal process}
In equilibrium renewal process ($\alpha > 1$), the Laplace transform of $H_2(t)$ with respect to $t$ yields 
\begin{equation}
\hat{H}_2(s) = \frac{2  [1 - \hat{\rho}_+(s)\hat{\rho}_-(s)^2 - \hat{\rho}_+(s)^2 \hat{\rho}_-(s) + \hat{\rho}_+(s) + \hat{\rho}_-(s) - \hat{\rho}_+(s)^2\hat{\rho}_-(s)^2 ]}{(\mu_+ + \mu_-) [1 - \hat{\rho}_+(s)\hat{\rho}_-(s)]^2 s^2}.
\end{equation}
The asymptotic behaviors are given by 
\begin{equation}
\hat{H}_2(s) = \left\{
\begin{array}{ll}
\dfrac{8}{(\mu_+ + \mu_-)^2 s^3} + \dfrac{4(\sigma_+^2 + \sigma_-^2)   }{(\mu_+ + \mu_-)^3 s^2}  + o(s^{-2}) \quad (2< \alpha ,~Case~3),\\
\\
\dfrac{8}{(\mu_+ + \mu_-)^2 s^3} + \dfrac{12a_+}{(\mu_+ + \mu_-)^3 s^{4-\alpha}}+ o(s^{\alpha-4}) \quad (1<\alpha < 2,~Case~2).
\end{array}
\right.
\end{equation}
In the long-time limit, the second moment of $N_t$ becomes 
\begin{equation}
H_2(t) = \left\{
\begin{array}{ll}
\dfrac{4t^2}{(\mu_+ + \mu_-)^2} + \dfrac{4(\sigma_+^2 + \sigma_-^2)  }{(\mu_+ + \mu_-)^3 }t +o(t) \quad (2< \alpha ,~Case~3),\\
\\
\dfrac{4t^2}{(\mu_+ + \mu_-)^2} + \dfrac{12a_+}{(\mu_+ + \mu_-)^3 \Gamma(4-\alpha)} t^{3-\alpha} +o(t^{3-\alpha}) \quad (1< \alpha <2,~Case~2).
\end{array}
\right.
\end{equation}
The variance of $N_t$ is given by
\begin{equation}
{\rm Var}(N_t)  \sim \left\{
\begin{array}{ll}
\dfrac{4(\sigma_+^2 + \sigma_-^2)  }{(\mu_+ + \mu_-)^3 }t  \quad (2<\alpha,~Case~3),\\
\\
\dfrac{12a_+ }{(\mu_+ + \mu_-)^3 \Gamma(4-\alpha)} t^{3-\alpha} \quad (2<\alpha ,~Case~2).
\end{array}
\right.
\end{equation}
The variance of $N_t$ increases as $t^{3-\alpha}$ for $1<\alpha <2$. However,  the coefficient of the variance
is different from that in the ordinary alternating renewal process. This phenomena is also observed in L\'evy walk model of superdiffusion 
\cite{Akimoto2012, Godec2013}. 

\subsubsection{aging alternating renewal process}

By a similar calculation of the first moment in the aging alternating renewal process, 
the asymptotic behavior of the double Laplace transform of the second moment of the number of renewals in $[t_a,t_a+t]$, i.e., 
$N_{t+t_a}-N_{t_a}$, with respect to $t$ and the aging time $t_a$ is approximately given by
\begin{equation}
\hat{H}_2(s; u)  \cong \frac{ 8\hat{f}_{E}(s;u)   }{s [1-\hat{\rho}_+(s) \hat{\rho}_-(s)]^2}.
\end{equation}
The leading order for Cases 1 and 4 becomes 
\begin{equation}
\hat{H}_2(s; u) \sim \left\{
\begin{array}{ll}
\dfrac{8}{a_+^2 s^{1+2\alpha} u}\quad (s\ll u),\\
\\
\dfrac{8}{a_+^2 s^{2+\alpha} u^{\alpha}} \quad (s \gg u).\\
\end{array}
\right.
\end{equation}
The inverse Laplace transform yields 
\begin{equation}
H_2(t; t_a) \sim \left\{
\begin{array}{ll}
\dfrac{8t^{2\alpha}}{a_+^2 \Gamma (1+ 2\alpha)}\quad (t \gg t_a,~Cases~1~and~4),\\
\\
\dfrac{8 t^{1+\alpha} t_a^{\alpha -1}}{a_+^2  \Gamma(\alpha)  \Gamma(2+\alpha)} \quad (t \ll t_a,~Cases~1~and~4).\\
\end{array}
\right.
\end{equation}

\subsection{asymptotic behaviors of  higher moments}
The $n$th moment of $N_t$ is also obtained as 
\begin{eqnarray}
H_n (t) \equiv \langle N_t ^n\rangle = \sum_{r=0}^\infty r^n \Pr (N_t=r)
=\sum_{r=1}^\infty \{ r^n - (r-1)^n \}  \Pr(t_r<t).
\end{eqnarray}
The asymptotic behavior of the Laplace transform with respect to $t$ for $s\to 0$ becomes  
\begin{equation}
\hat{H}_n (s) \sim  \frac{n}{s}\sum_{r=1}^\infty r^{n-1}  \hat{\rho}_1(s) \cdots \hat{\rho}_r(s).
\end{equation}

\subsubsection{ordinary alternating renewal process}
In the ordinary renewal process, we have
\begin{equation}
\hat{H}_n (s) \sim  \frac{2^n n}{s} \sum_{r=1}^\infty  r^{n-1} \{ \hat{\rho}_+(s) \hat{\rho}_-(s) \}^r 
\sim \frac{2^n n!}{s [ 1 - \hat{\rho}_+(s) \hat{\rho}_-(s)]^{n}}.
\label{ordinary Hn}
\end{equation}
The asymptotic behaviors are given by 
\begin{equation}
\hat{H}_n(s) = \left\{
\begin{array}{ll}
\dfrac{2^n n!}{(\mu_+ + \mu_-)^{n} s^{n+1}}  + o(s^{-n-1}) \quad (1< \alpha ,~Cases~2~and~3),\\
\\
\dfrac{2^n n!}{a_+^n s^{n\alpha+1}}  + o(s^{-n\alpha-1}) \quad (\alpha_+<\alpha_-<1,~Cases~1~and~4).\\
\end{array}
\right.
\end{equation}
In the long-time limit, the $n$th moment of $N_t$ becomes 
\begin{equation}
H_n(t) = \left\{
\begin{array}{ll}
 \left( \dfrac{t}{\mu}\right)^n  \quad (1< \alpha ,~Cases~1~and~4),\\
\\
\dfrac{2^n n!}{a_+^n \Gamma(n\alpha+1)} t^{n\alpha} \quad (\alpha_+<\alpha_-<1,~Cases~1~and~4).
\end{array}
\right.
\label{higher moments Nt ord}
\end{equation}
\end{widetext}

\subsubsection{equilibrium alternating renewal process}
In equilibrium renewal process ($\alpha > 1$), the asymptotic behaviors of the higher moments of $N_t$ are the same as Eq.~(\ref{ordinary Hn}). 
Thus, for Cases 2 and 3, we have 
\begin{equation}
H_n(t) \sim \left( \dfrac{t}{\mu}\right)^n .
\end{equation}

\subsubsection{aging alternating renewal process}

For Cases 1 and 4, the asymptotic behavior of the double Laplace transform of the number of renewals in $[t_a,t_a+t]$, i.e., 
$N_{t+t_a}-N_{t_a}$, with respect to $t$ and the aging time $t_a$ is approximately given by
\begin{equation}
\hat{H}_n (s; u)  \cong \frac{ 2^n n!\hat{f}_{E}(s;u)   }{s [1-\hat{\rho}_+(s) \hat{\rho}_-(s)]^n}.
\end{equation}
The leading order becomes 
\begin{equation}
\hat{H}_n(s; u) \sim \left\{
\begin{array}{ll}
\dfrac{2^n n!}{a_+^n s^{1+n \alpha} u}\quad (s\ll u),\\
\\
\dfrac{2^n n!}{a_+^n s^{2+ n\alpha} u^{\alpha}} \quad (s \gg u).\\
\end{array}
\right.
\end{equation}
The inverse Laplace transform yields 
\begin{equation}
H_n(t; t_a) \sim \left\{
\begin{array}{ll}
\dfrac{2^n n! t^{n\alpha}}{a_+^n \Gamma (1+ n\alpha)}\quad (t \gg t_a),\\
\\
\dfrac{2^n n! t^{1+ n\alpha} t_a^{\alpha -1}}{a_+^n  \Gamma(\alpha)  \Gamma(2+n\alpha)} \quad (t \ll t_a).\\
\end{array}
\right.
\end{equation}

\section{Occupation time statistics}

Here, we consider the distribution of occupation times $T_t^+$ and the moments of $T_t^+$ as a function of time $t$. 
We define the joint probability distribution of $T_t^+=y$ and $N_t=n$ with $\sigma (0)=\pm 1$ as 
\begin{equation}
g_n^{\pm}(y;t) = \langle \delta(y-T_t^+) I(t_n\leq t<t_{n+1}\rangle.
\end{equation}
The double Laplace transform of $g_n^{\pm}(y;t)$ with respect to $y$ and $t$ is given by
\begin{widetext}
\begin{eqnarray}
\hat{g}_n^{\pm}(u;s) = \int_0^\infty dt e^{-st} \int_0^\infty dy e^{-uy} \langle \delta(y-T_t^+) I(t_n\leq t<t_{n+1}\rangle
= \left\langle \int_{t_n}^{t_{n+1}} dt e^{-st} e^{-u T_t^+} \right\rangle.
\end{eqnarray}
For $\sigma (0)=+1$, we have
\begin{eqnarray}
\hat{g}_{2k+1}^{+}(u;s) 
=\frac{1 - \hat{\rho}_-(s)}{s} \hat{\rho}_-^k(s) \hat{\rho}_+^{k}(s+u) \hat{\rho}_1(s+u).
\label{pdf_+_2k+1}
\end{eqnarray}
and
\begin{eqnarray}
\hat{g}_{2k}^{+}(u;s) 
&=& \left\{
\begin{array}{ll}
\dfrac{1- \hat{\rho}_+(s+u)}{s+u} \hat{\rho}_-^k(s) \hat{\rho}_+^{k-1}(s+u)\hat{\rho}_1(s+u) \quad &(k\geq 1),\\
\\
\dfrac{1 - \hat{\rho}_1(s+u)}{s+u} \quad &(k=0).
\end{array}
\right.
\label{pdf_+_2k}
\end{eqnarray}
For $\sigma_0=-1$, we have 
\begin{eqnarray}
\hat{g}_{2k}^{-}(u;s) 
= \left\{
\begin{array}{ll}
\dfrac{1- \hat{\rho}_-(s)}{s} \hat{\rho}_1(s) \hat{\rho}_-^{k-1}(s) \hat{\rho}_+^k(s+u) \quad &(k\geq 1),\\
\\
\dfrac{1- \hat{\rho}_1(s)}{s} &(k=0).
\end{array}
\right.
\label{pdf_-_2k}
\end{eqnarray}
and
\begin{eqnarray}
\hat{g}_{2k+1}^{-}(u;s) 
&=& \frac{1- \hat{\rho}_+(s+u)}{s+u} \hat{\rho}_+^k(s+u) \hat{\rho}_-^{k}(s) \hat{\rho}_1(s).
\label{pdf_-_2k+1}
\end{eqnarray}

\subsection{Fluctuations of $T_t^+/t$}

\subsubsection{ordinary alternating renewal process}
Here, we consider the distribution of $T_t^+/t$ for an ordinary alternating renewal process. 
Using Eqs. (\ref{pdf_+_2k+1}), (\ref{pdf_+_2k}), (\ref{pdf_-_2k}), (\ref{pdf_-_2k+1}), we have the Laplace transform of the PDF of $T_t^+$:
\begin{eqnarray}
\hat{g}^+(u;s) 
&=& \sum_{n=0}^\infty \{\hat{g}_{2n}^+(u;s) + \hat{g}_{2n+1}^+(u;s)\}  
= \frac{1}{1- \hat{\rho}_+(s+u)\hat{\rho}_-(s)}  \left\{ \frac{1-\hat{\rho}_-(s)}{s} \hat{\rho}_+(s+u) + \frac{1-\hat{\rho}_+(s+u)}{s+u} \right\}
\label{gus+ ordinary}
\end{eqnarray}
and
\begin{eqnarray}
\hat{g}^-(u;s) 
&=& \sum_{n=0}^\infty \{\hat{g}_{2n}^- (u;s) + \hat{g}_{2n+1}^- (u;s)\}  
= \frac{1}{1- \hat{\rho}_+(s+u)\hat{\rho}_-(s)}  \left\{ \frac{1-\hat{\rho}_+(s+u)}{s+u} \hat{\rho}_-(s)+ \frac{1-\hat{\rho}_-(s)}{s}  \right\}.
\end{eqnarray}
In the small $s$ and $u$ limit, 
\begin{equation}
\hat{g}^+(u;s)  \sim \hat{g}^-(u;s) 
\sim \frac{1}{1- \hat{\rho}_+(s+u)\hat{\rho}_-(s)}  \left\{ \frac{1-\hat{\rho}_-(s)}{s}  + \frac{1-\hat{\rho}_+(s+u)}{s+u} \right\}.
\label{gus ordinary}
\end{equation}

For $\alpha=\alpha_+=\alpha_- <1$ (Case 1), fluctuations of $x=T_t^+/t$ are intrinsic even in the long-time limit. The double Laplace transform
for $\hat{g}^\pm (u;s)$ becomes 
\begin{equation}
\hat{g}^+(u;s)  \sim \hat{g}^-(u;s) 
\sim \frac{a_+(s+u)^{\alpha-1}+a_-s^{\alpha-1}}{a_+(s+u)^{\alpha} +a_-s^{\alpha}}
=\frac{1}{s} \frac{a_+(1+u/s)^{\alpha-1}+a_-}{a_+(1+u/s)^{\alpha} +a_-}.
\end{equation}
By the method of the inverse Laplace transform given in Appendix~B in \cite{God2001}, the PDF $\tilde{g}(x)$ of $x=T_t^+/t$ for  $\alpha=\alpha_+=\alpha_- <1$ (Case 1)
in the long-time limit becomes 
\begin{equation}
\lim_{t\to\infty} \tilde{g}(x) = \frac{a \sin \pi \alpha}{\pi} 
\frac{x^{\alpha-1} (1-x)^{\alpha-1}}{a^2 x^{2\alpha} +2a \cos \pi \alpha (1-x)^\alpha x^\alpha + (1-x)^{2\alpha}},
\end{equation}
where $a=a_-/a_+$. 
This is the exactly same as the Lamperti's generalized arcsine law \cite{Lamperti1958}. Counter-intuitively, the ratio of occupation time in the positive side
 $T_t^+/t$ does not converge to a constant but remains random for $\alpha<1$ even in the long-time limit. For  $\alpha>1$ (Cases 2 and 3), on the other hand, 
 $T_t^+/t$ converges to $\mu_+/(\mu_+ + \mu_-)$ for $t\to\infty$. 

\subsubsection{equilibrium alternating renewal process}
We consider fluctuations of $T_t^+$ in an equilibrium alternating renewal process for $\alpha>1$ (Cases 2 and 3). For $N_t=2k+1$ ($k=0, 1, \cdots$) with $\sigma_0=+1$, we have
\begin{eqnarray}
\hat{g}_{2k+1}^{+}(u;s) &=& \frac{1- \hat{\rho}_-(s)}{s} \hat{\rho}_-^k(s) \hat{\rho}_+^{k}(s+u) \hat{f}_+(s+u),
\end{eqnarray}
where $\hat{f}_+(s)$ is the Laplace transform of $f_+(x)$.
For $N_t=2k$ ($k=0, 1, \cdots$) with $\sigma_0=+1$, we have
\begin{eqnarray}
\hat{g}_{2k}^{+}(u;s) 
&=& \left\{
\begin{array}{ll}
 \dfrac{1- \hat{\rho}_+(s+u)}{s+u} \hat{\rho}_-^k(s) \hat{\rho}_+^{k-1}(s+u) \hat{f}_+(s+u) \quad &(k\geq 1),\\
\\
\dfrac{1-\hat{f}_+(s+u)}{s+u} &(k=0).
\end{array}
\right.
\end{eqnarray}
Moreover, we have
\begin{equation}
\hat{g}_{2k+1}^{-}(u;s) =  \frac{1- \hat{\rho}_+(s+u)}{s+u} \hat{f}_-(s) \hat{\rho}_+^{k}(s+u) \hat{\rho}_-^{k}(s),
\end{equation}
and
\begin{equation}
\hat{g}_{2k}^{-}(u;s) = \left\{
\begin{array}{ll}
 \dfrac{1- \hat{\rho}_-(s)}{s} \hat{f}_-(s) \hat{\rho}_+^{k}(s+u) \hat{\rho}_-^{k-1}(s) \quad &(k\geq 1),\\
\\
 \dfrac{1-\hat{f}_-(s)}{s} &(k=0),
\end{array}
\right.
\end{equation}
where $\hat{f}_-(s)$ is the Laplace transform of $f_-(x)$. 
It follows that the Laplace transform of the PDF of $T_t^+$ is given by
\begin{eqnarray}
\hat{g}(u;s) &=&  P_+ \hat{g}^+(u;s) + P_- \hat{g}^-(u;s) \\
&=& P_+ \left\{ \frac{1- \hat{f}_+(s+u)}{s+u}+\frac{\hat{f}_+(s+u)}{1- \hat{\rho}_+(s+u)\hat{\rho}_-(s)} \left[ \frac{1-\hat{\rho}_+(s+u)}{s+u}  
\hat{\rho}_-(s) +\frac{1-\hat{\rho}_-(s)}{s}\right] \right\}\nonumber\\
&&+ P_- \left\{ \frac{1-\hat{f}_-(s)}{s} + \frac{\hat{f}_-(s)}{1- \hat{\rho}_+(s+u)\hat{\rho}_-(s)} \left[ \frac{1-\hat{\rho}_+(s+u)}{s+u}  
+\frac{1-\hat{\rho}_-(s)}{s}\hat{\rho}_+(s+u)\right] \right\}.
\label{gus eq}
\end{eqnarray}
For $\alpha>1$ (Cases 2 and 3), the Laplace transform of the PDF of $T_t^+$ in the asymptotic limit ($s<u \ll 1$) becomes 
\begin{eqnarray}
\hat{g}(u;s) \sim  \frac{1}{1- \hat{\rho}_+(s+u)\hat{\rho}_-(s)} \left[ \frac{1-\hat{\rho}_+(s+u)}{s+u}  
 +\frac{1-\hat{\rho}_-(s)}{s}\right] ,
\end{eqnarray}
which is the same as that for the ordinary alternating renewal process.

\subsubsection{aging alternating renewal process}
Here, we consider a particular case of $\alpha=\alpha_-=\alpha_+$ (Case 1) and occupation time of $+$ state in $[t_a, t_a+t]$ denoted by 
$T^+_{t_a,t}$.
Because there is no equilibrium distribution for the forward recurrence time in Case 1, the occupation time intrinsically depends on 
time $t_a$ ({\it aging}). This aging extension of the generalized arcsine law was established in Ref.~\cite{Akimoto2020}. 
The PDF  of $y=T_{t_a,t}^+$ denoted by $g(y; t,t_a)$ is given by
\begin{equation}
g(y; t,t_a) = P_+(t_a) g^+ (y; t,t_a) + P_-(t_a) g^-(y; t,t_a). 
\end{equation}
The Laplace transform of $g(y; t,t_a)$ with respect to $y,t$, and $t_a$  
($y \leftrightarrow u, t \leftrightarrow s$ and $t_a \leftrightarrow v$)  becomes 
\begin{eqnarray}
\hat{g}(u;s,v) &=& \int_0^\infty dt e^{-st} \int_0^\infty dt_a e^{-vt_a} \int_0^\infty dy e^{-uy} \langle \delta(y-T_{t_a,t}^+) \rangle\\
&=&  \frac{\hat{f}_{E,+}(0,v)- \hat{f}_{E,+}(s+u,v)}{s+u}+\frac{\hat{f}_{E,+}(s+u,v)}{1- \hat{\rho}_+(s+u)\hat{\rho}_-(s)} \left[ \frac{1-\hat{\rho}_+(s+u)}{s+u}  
\hat{\rho}_-(s) +\frac{1-\hat{\rho}_-(s)}{s}\right] \nonumber\\
&&+  \frac{\hat{f}_{E,+}(0,v)-\hat{f}_{E,-}(s,v)}{s} + \frac{\hat{f}_{E,-}(s,v)}{1- \hat{\rho}_+(s+u)\hat{\rho}_-(s)} \left[ \frac{1-\hat{\rho}_+(s+u)}{s+u}  
+\frac{1-\hat{\rho}_-(s)}{s}\hat{\rho}_+(s+u)\right] .
\end{eqnarray}
In the asymptotic limit ($u, s \ll 1$), we have
\begin{eqnarray}
\hat{g}(u;s,v) \sim  
\frac{ \hat{f}_{E,+}(s+u,v) +  \hat{f}_{E,-}(s,v)}{1- \hat{\rho}_+(s+u)\hat{\rho}_-(s)} \left[ \frac{1-\hat{\rho}_+(s+u)}{s+u}  
 +\frac{1-\hat{\rho}_-(s)}{s}\right].
 \label{gusv asympt aging}
\end{eqnarray}
For $u,s \ll v$, we obtain $\hat{g}(u;s,v) \sim \hat{g}(u;s)/v$. Therefore, there is no explicit dependence of  the distribution on $t_a$ 
for $u,s \ll v$, i.e., $t_a \ll t, y$. 
On the other hand, there is an explicit dependence of  the distribution on $t_a$ for $u,s \gg v$, i.e., $t_a \gg t$. 

\subsection{first moment of $T_t^+$}
Here, we consider the first moment of $T_t^+$. The Laplace transform of the first moment of $T_t^+$ with respect to $t$ is defined as
\begin{eqnarray}
\hat{T}_1(s) \equiv \int_0^\infty e^{-st}  \langle T_t^+ \rangle dt.
\end{eqnarray}

\subsubsection{ordinary alternating renewal process}
The Laplace transform of the first moment of $T_t^+$ with respect to $t$ is obtained from Eq.~(\ref{gus+ ordinary}):
\begin{eqnarray}
\hat{T}_1(s) = - \left. \frac{\partial \hat{g}^+(u;s)}{\partial u}\right|_{u=0}.
\end{eqnarray}
The asymptotic behaviors of $\hat{T}_1(s)$ are given by 
\begin{equation}
\hat{T}_1(s) = \left\{
\begin{array}{ll}
\dfrac{\mu_+}{(\mu_++\mu_-)s^2}- \dfrac{\sigma_+^2\mu_- - \sigma_-^2 \mu_+ -  \mu_+\mu_- (\mu_++\mu_-)   }{2(\mu_+ + \mu_-)^2 s}  + o(s^{-1}) \quad (2< \alpha,~Case~3),\\
\\
\dfrac{\mu_+}{(\mu_++\mu_-)s^2} - \dfrac{a_+ \mu_- }{(\mu_+ + \mu_-)^2 s^{3-\alpha}}+
 o(s^{\alpha-3}) \quad (1<\alpha < 2,~Case~2),\\
\\
\dfrac{1}{s^2} + o(s^{-2}) \quad (\alpha < 1,~Cases~1~and~4).
\end{array}
\right.
\end{equation}
Therefore, in the long-time limit, we have
\begin{equation}
\langle T_t^+ \rangle = \left\{
\begin{array}{ll}
 \dfrac{\mu_+}{\mu_++\mu_-} t - \dfrac{\sigma_+^2\mu_- - \sigma_-^2 \mu_+ -  \mu_+\mu_- (\mu_++\mu_-)   }{2(\mu_+ + \mu_-)^2 }  \quad (2< \alpha,~Case~3),\\
\\
\dfrac{\mu_+}{\mu_++\mu_-} t - \dfrac{a_+ \mu_- }{\Gamma(3-\alpha) (\mu_+ + \mu_-)^2} t^{2-\alpha} + o(t^{2-\alpha}) \quad (1<\alpha < 2,~Case~2),\\
\\
t+ o(t) \quad (\alpha < 1,~Cases~1~and~4).
\end{array}
\right.
\label{occupation time 1st ord}
\end{equation}

\subsubsection{equilibrium alternating renewal process}

The Laplace transform of the first moment of $T_t^+$ in an equilibrium alternating renewal process is obtained from Eq.~(\ref{gus eq}): 
\begin{eqnarray}
\hat{T}_1(s) =  \frac{\mu_+}{(\mu_++\mu_-)s^2}.
\end{eqnarray}
Therefore, for Cases 2 and 3, we have in the equilibrium alternating renewal process 
\begin{equation}
\langle T_t^+ \rangle = \frac{\mu_+}{\mu_++\mu_-} t.
\end{equation}

\subsubsection{aging alternating renewal process}

The asymptotic behavior of $\hat{T}_1(s) $ in an aging alternating renewal process $(\alpha<1)$ is obtained from Eq.~(\ref{gusv asympt aging}).
For $u,s \ll v$, the result is the same as that for the ordinary alternating renewal process, i.e., $\langle T_t^+ \rangle \sim t$.  Moreover,  for $u,s \gg v$, i.e., $t\ll t_a$, 
we have the same result, i.e., $\langle T_t^+ \rangle \sim t$.

\subsection{second moment}

Here, we consider the second moment of $T_t^+$. The Laplace transform of the second moment of $T_t^+$ with respect to $t$ is defined as
\begin{eqnarray}
\hat{T}_2(s) \equiv \int_0^\infty e^{-st}  \langle (T_t^+)^2 \rangle dt.
\end{eqnarray}

\subsubsection{ordinary alternating renewal process}
The Laplace transform of the second moment of $T_t^+$ with respect to $t$ is obtained from 
\begin{eqnarray}
\hat{T}_2(s) =  \left. \frac{\partial^2 \hat{g}^+(u;s)}{\partial u^2}\right|_{u=0}.
\end{eqnarray}
The asymptotic behaviors of $\hat{T}_2(s)$ are given by 
\begin{equation}
\hat{T}_2(s) = \left\{
\begin{array}{ll}
\dfrac{2\mu_+^2}{(\mu_++\mu_-)^2s^3}+ \dfrac{\sigma_+^2 \mu_- (\mu_- - \mu_+) + 2 \sigma_-^2 \mu_+^2  +  \mu_+^2\mu_- (\mu_++5\mu_-)  }{(\mu_+ + \mu_-)^3 s^2}  + o(s^{-1}) \quad (2< \alpha ,~Case~3),\\
\\
\dfrac{2\mu_+^2}{(\mu_++\mu_-)^2s^3} - \dfrac{2 a_+ \mu_- ( (3-\alpha) \mu_+ - (\alpha -1) \mu_-)}{(\mu_+ + \mu_-)^3 s^{4-\alpha}}+
 o(s^{\alpha-4}) \quad (1<\alpha < 2,~Case~2).
\end{array}
\right.
\end{equation}
Therefore, in the long-time limit, 
\begin{equation}
\langle (T_t^+)^2 \rangle = \left\{
\begin{array}{ll}
\left(\dfrac{\mu_+}{\mu_++\mu_-} t \right)^2+ \dfrac{\sigma_+^2 \mu_- (\mu_- - \mu_+) + 2 \sigma_-^2 \mu_+^2  +  \mu_+^2\mu_- (\mu_++5\mu_-) }{(\mu_+ + \mu_-)^3 } t \quad (2< \alpha ,~Case~3),\\
\\
\left(\dfrac{\mu_+}{\mu_++\mu_-} t \right)^2 - \dfrac{4 a_+ \mu_+\mu_- }{\Gamma(4-\alpha) (\mu_+ + \mu_-)^3} t^{3-\alpha} + o(t^{3-\alpha}) \quad (1<\alpha < 2,~Case~2).
\end{array}
\right.
\label{second occupation time ord}
\end{equation}
It follows that the relative standard deviation of $T_t^+$ is given by
\begin{equation}
\frac{\sqrt{\langle (T_t^+)^2 \rangle - \langle T_t^+ \rangle^2}}{\langle T_t^+ \rangle} \sim  \left\{
\begin{array}{ll}
\sqrt{ \dfrac{ 4 \mu_-^2 + \sigma_-^2  }{\mu_+ + \mu_- } } t^{-\frac{1}{2}}\quad (2< \alpha ,~Case~3),\\
\\
\sqrt{\dfrac{2 a_+ \mu^2 }{\Gamma(4-\alpha) (\mu_+ + \mu_-)}} t^{- \frac{\alpha-1}{2}}  \quad (1<\alpha < 2,~Case~2).
\end{array}
\right.
\label{T RSD cases23}
\end{equation}
where $\mu= \mu_-/\mu_+$. For $1<\alpha<2$, the relaxation becomes slower than that for $\alpha>2$. This slow relaxation is observed 
in diffusion of lipid molecules \cite{Akimoto2011}.

\subsubsection{equilibrium alternating renewal process}

The asymptotic behaviors of $\hat{T}_2(s)$ are given by 
\begin{equation}
\hat{T}_2(s) = \left\{
\begin{array}{ll}
\dfrac{2\mu_+^2}{(\mu_++\mu_-)^2s^3}+ \dfrac{ 4P_-\mu_+\mu_- (\sigma_+^2  + \mu_+^2)  +  2\mu_+^2(\mu_+\mu_-   +\sigma_-^2)  }{(\mu_+ + \mu_-)^3 s^2}  + o(s^{-1}) \quad (2< \alpha ,~Case~3),\\
\\
\dfrac{2\mu_+^2}{(\mu_++\mu_-)^2s^3} - \dfrac{(3-\alpha) a_+ \mu_-^2}{(\mu_+ + \mu_-)^3 s^{4-\alpha}}+
 o(s^{\alpha-4}) \quad (1<\alpha < 2,~Case~2).
\end{array}
\right.
\end{equation}
Therefore, in the long-time limit, 
\begin{equation}
\langle (T_t^+)^2 \rangle = \left\{
\begin{array}{ll}
\left(\dfrac{\mu_+}{\mu_++\mu_-} t \right)^2+ \dfrac{ 4P_-\mu_+\mu_- (\sigma_+^2  + \mu_+^2)  +  2\mu_+^2(\mu_+\mu_-   +\sigma_-^2)  }{(\mu_+ + \mu_-)^3 }  t \quad (2< \alpha ,~Case~3),\\
\\
\left(\dfrac{\mu_+}{\mu_++\mu_-} t \right)^2 - \dfrac{a_+ \mu_-^2 }{\Gamma(3-\alpha) (\mu_+ + \mu_-)^3} t^{3-\alpha} + o(t^{3-\alpha}) \quad (1<\alpha < 2,~Case~2).
\end{array}
\right.
\end{equation}
It follows that the relative standard deviation of $T_t^+$ is given by
\begin{equation}
\frac{\sqrt{\langle (T_t^+)^2 \rangle - \langle T_t^+ \rangle^2}}{\langle T_t^+ \rangle} \sim  \left\{
\begin{array}{ll}
\sqrt{ \dfrac{ 4P_-\mu (\sigma_+^2  + \mu_+^2)  +  2(\mu_+\mu_-   +\sigma_-^2)  }{\mu_+ + \mu_-}  } t^{-\frac{1}{2}}\quad (2< \alpha ,~Case~3),\\
\\
\sqrt{\dfrac{ a_+ \mu^2 }{\Gamma(3-\alpha) (\mu_+ + \mu_-)}} t^{- \frac{\alpha-1}{2}}  \quad (1<\alpha < 2,~Case~2).
\end{array}
\right.
\label{T RSD cases23 eq}
\end{equation}
\end{widetext}

\section{ergodic properties}
Here, we discuss ergodic properties for alternating renewal processes. Time average of an observable $f(x)$ in an alternating renewal process 
is defined by
\begin{equation}
\overline{f(t)} \equiv \frac{1}{t} \int_0^t f(\sigma(t')) dt'.
\end{equation}
If the system is ergodic, time averages converge to the ensemble average in the long-time limit:
\begin{equation}
\overline{f(t)} \to \langle f(x) \rangle_{\rm eq} \quad {\rm for}~t\to \infty
\label{ergodic Nt}
\end{equation}
for all trajectories $\sigma(t)$, where $\langle \cdot \rangle_{\rm eq}$ is the equilibrium ensemble average. 

\subsection{the number of renewals}
 We consider the time average of the number of renewals, i.e., 
$N_t/t$. If the system is ergodic, 
\begin{equation}
\frac{N_t}{t} \to \lambda_{\rm eq} \quad {\rm for}~t\to \infty
\label{ergodic Nt}
\end{equation}
for all paths $N_t$, where $\lambda_{\rm eq}$ is the equilibrium jump rate. 
For $\alpha>1$ (Cases 2 and 3),
by Eq.~(\ref{renewal function}), we have 
$\left\langle N_t/t \right\rangle \to 1/\mu$ for $t \to \infty$. Moreover, by Eq.~(\ref{variance Nt ord}), the variance of $N_t/t$ becomes zero in the long-time limit:
$\langle \left( N_t/t \right)^2\rangle  - \left\langle  N_t/t \right\rangle^2 \to 0$ for $t \to \infty$. 
Therefore,  alternating renewal processes with $\alpha>1$ are ergodic in the sense that Eq.~(\ref{ergodic Nt}) holds with $\lambda_{\rm eq}=1/\mu$. 

For $\alpha<1$ (Cases 1 and 4), the ergodic properties become different from those for $\alpha>1$. By Eq.~(\ref{renewal function}), we have 
\begin{equation}
\left\langle N_t/t^\alpha \right\rangle \to \frac{2}{ a_+ \Gamma(1+\alpha)}
\end{equation}
 for $t \to \infty$. Moreover, by Eq.~(\ref{variance Nt ord}), the variance of $N_t/t$ becomes zero in the long-time limit:
\begin{equation}
\langle \left( N_t/t^\alpha \right)^2\rangle  - \left\langle  N_t/t^\alpha \right\rangle^2 \to \dfrac{4(2 \Gamma(\alpha+1)^2 - \Gamma(2\alpha +1))} {a_+^2 \Gamma(2\alpha +1)\Gamma(\alpha+1)^2}
\end{equation}
 for $t \to \infty$. Therefore, the variance of $N_t/t^\alpha$ is a non-zero constant. In other words, $N_t/t^\alpha$ does not converge to a constant even in the long-time 
 limit but remains random. Therefore, alternating renewal processes with $\alpha<1$ are not ergodic. 
 Using the results for the higher moments, i.e., Eq.~(\ref{higher moments Nt ord}),  we have the asymptotic behavior of the
  $n$th moment of $N_t /\langle N_t \rangle$.
In particular, the $n$th moment of $N_t /\langle N_t \rangle$ converges to $n! \Gamma (1+\alpha)^n/ \Gamma(n\alpha+1)$ for $t\to\infty$. This is the $n$th moment 
of the Mittag-Leffler distribution. In other words, $N_t /\langle N_t \rangle$ shows trajectory-to-trajectory fluctuations and 
the distribution of $N_t /\langle N_t \rangle$ converges to the Mittag-Leffler distribution. This distribution appears in 
stochastic processes \cite{Darling1957, Miyaguchi2013, Metzler2014,AkimotoYamamoto2016a} as well as the infinite ergodic theory \cite{Akimoto2010, Aaronson1997}. 

\subsection{occupation times}
 We consider the time average of an occupation time, i.e., 
$f(x)=I_{(0,\infty)}(x)$, where $I_A(x)$ is the indicator function of $A$. If the system is ergodic, 
\begin{equation}
\frac{1}{t} \int_0^{t} I_{(0,\infty)} (\sigma(t')) dt' \to \langle I_A(x) \rangle_{\rm eq} \quad {\rm for}~t\to \infty
\label{ergodic Nt}
\end{equation}
for all trajectories $\sigma(t)$. The left-hand side is the ratio of the occupation time, i.e., $T_t^+/t$.  
For $\alpha>1$ (Cases 2 and 3),
by Eq.~(\ref{occupation time 1st ord}), we have 
$\left\langle T_t^+/t \right\rangle \to \mu_+/(\mu_+ + \mu_-)$ for $t \to \infty$. Moreover, by Eq.~(\ref{second occupation time ord}), 
the variance of $T_t^+/t$ becomes zero in the long-time limit:
$\langle \left( T_t^+/t \right)^2\rangle  - \left\langle  T_t^+/t \right\rangle^2 \to 0$ for $t \to \infty$. 
Therefore,  alternating renewal processes with $\alpha>1$ are ergodic in the sense that $T_t^+/t$ converges to $\mu_+/(\mu_+ + \mu_-)$ 
in the long-time limit. Although alternating renewal processes with $\alpha>1$ are ergodic, alternating renewal processes with $1<\alpha<2$ (Case 2)  
exhibits a slow relaxation, i.e., $\langle \left( T_t^+/t \right)^2\rangle  - \left\langle  T_t^+/t \right\rangle^2 \propto t^{-\frac{\alpha-1}{2}}$ for $t \to \infty$. 

For $\alpha_+=\alpha_-<1$ (Case 1), the ergodic properties are completely different from those for $\alpha>1$. As shown in section VA.1, $T_t^+/t $ 
does not converge to a constant but exhibits trajectory-to-trajectory fluctuations. The distribution of $T_t^+/t $ converges to the generalized 
arcsine distribution in the long-time limit.  Therefore, ergodicity of the alternating renewal process breaks down. We note 
$T_t^+/t  \to 1$ for $t\to \infty$ for $\alpha_+ < \alpha_- < 1$ (Case 1) and $\alpha<1, \alpha_->1$ (Case 4). 

\section{Correlation function}
We consider the correlation function of $D(t)$, which is defined by $D(t)=D_+$ or $D_-$ when $\sigma(t)=+1$ or $\sigma(t)=+1$, respectively. 
If there exists an equilibrium distribution for $D(t)$, the correlation function is defined by
\begin{eqnarray}
C(t) \equiv \langle D(0) D(t) \rangle - \langle D(0) \rangle\langle D \rangle_{\rm eq},
\end{eqnarray}
where $ \langle D \rangle_{\rm eq}$ is the equilibrium ensemble average of $D$, which is given by
\begin{equation}
\langle D \rangle_{\rm eq} = \frac{D_+\mu_+ +D_- \mu_-}{\mu_+ + \mu_-}.
\label{average D eq} 
\end{equation}

\subsubsection{ordinary alternating renewal process}

In the ordinary alternating renewal process, if there exists an equilibrium distribution, the correlation function is represented as 
\begin{eqnarray}
C(t) = D_+^2 W_{++}(t) + D_+D_- W_{+-}(t) - D_+ \langle D \rangle_{\rm eq}, 
\end{eqnarray}
where $W_{++}(t) = \Pr \{D(t)=D_+| D(0) =D_+\}$ and $W_{+-}(t) = \Pr \{D(t)=D_-| D(0) =D_+\}$. The conditional probabilities are written as
\begin{widetext}
\begin{eqnarray}
W_{++}(t) &=& \sum_{n=0}^\infty \Pr (N_t=2n) = \sum_{n=1}^\infty \{\Pr(N_t < 2n+1) - \Pr (N_t<2n)\} + \Pr(N_t=0)\\
&=& \sum_{n=1}^\infty \{\Pr(S_{2n} < t) - \Pr (S_{2n+1}<t)\} + \Pr(N_t=0),
\end{eqnarray}
and
\begin{eqnarray}
W_{+-}(t) &=& \sum_{n=0}^\infty \Pr (N_t=2n+1) = \sum_{n=0}^\infty \{\Pr(N_t < 2n+2) - \Pr (N_t<2n+1)\} \\
&=& \sum_{n=0}^\infty \{\Pr(S_{2n+1} < t) - \Pr (S_{2n+2}<t)\} .
\end{eqnarray}
The Laplace transforms are given by
\begin{eqnarray}
\hat{W}_{++}(s)&=& \frac{1}{s} \sum_{n=1}^\infty \hat{f}_{E,+}(s) \{\hat{\rho}_-(s)\hat{\rho}_+(s)\}^{n-1} \hat{\rho}_-(s)
- \frac{1}{s} \sum_{n=1}^\infty \hat{f}_{E,+}(s) \{\hat{\rho}_-(s)\hat{\rho}_+(s)\}^{n}
+\frac{1-\hat{f}_{E,+}(s)}{s}\\
&=&\frac{1}{s} - \frac{\hat{\rho}_{+}(s)}{s} \frac{1-\hat{\rho}_-(s)}{1-\hat{\rho}_+(s)\hat{\rho}_-(s)}.
\end{eqnarray}
\end{widetext}
and
\begin{equation}
\hat{W}_{+-}(s)=\frac{\hat{\rho}_{+}(s)}{s} \frac{1-\hat{\rho}_-(s)}{1 -\hat{\rho}_+(s) \hat{\rho}_-(s)},
\end{equation}
It follows that the Laplace transform of $C(t)$ is given by
\begin{equation}
\hat{C}(s)= \frac{\mu_- D_+(D_+ - D_-) }{(\mu_+ + \mu_-)s} 
-  \frac{ D_+(D_+-D_-) \hat{\rho}_+(s) \{1-\hat{\rho}_-(s)\}}{\{1-\hat{\rho}_+(s)\hat{\rho}_-(s)\}s}.
\end{equation}

When both of the duration-time PDFs follow exponential distributions, we have
\begin{equation}
\hat{C}(s) = \frac{D_+ (D_+ - D_-)\mu_-}{\mu_+ + \mu_-} \frac{1}{(\mu_++\mu_-)/(\mu_+\mu_-) + s}.
\end{equation}
The inverse Laplace transform yields 
\begin{equation}
C(t) = \frac{D_+ (D_+ - D_-) \mu_-}{\mu_+ + \mu_-} \exp \left(-\frac{\mu_++\mu_-}{\mu_+\mu_-}t\right).
\end{equation}
For $1<\alpha<2$ (Case 2), we have 
\begin{equation}
\hat{C}(s) \sim - \frac{ a_+ \mu_- D_+ (D_+ - D_-) }{(\mu_+ + \mu_-)^2  s^{2 - \alpha}}
\end{equation}
for $s \to 0$. The inverse Laplace transform reads
\begin{equation}
C(t) \sim - \frac{a_+ \mu_- D_+ (D_+ - D_-) }{\Gamma(2-\alpha)(\mu_+ + \mu_-)^2} t^{-(\alpha-1)}
 \quad (t\to\infty). 
\end{equation}

\subsubsection{equilibrium alternating renewal process}

In equilibrium alternating renewal process, the correlation function is represented as 
\begin{widetext}
\begin{eqnarray}
C(t) = D_+^2 P_+ W_{++}(t) + D_+D_- P_+ W_{+-}(t)
+ D_-D_+ P_- W_{-+} + D_-^2 P_- W_{--}(t) - \langle D(0) \rangle^2, 
\end{eqnarray}
where $W_{-+}(t) = \Pr \{D(t)=D_+| D(0) =D_-\}$ and $W_{--}(t) = \Pr \{D(t)=D_-| D(0) =D_-\}$. 
The Laplace transforms of $W_{--}(t)$ and $W_{-+}(t)$ are given by
\begin{equation}
\hat{W}_{--}(s)=\frac{1}{s} - \frac{\hat{f}_{E,-}(s)}{s} \frac{1-\hat{\rho}_+(s)}{1-\hat{\rho}_+(s)\hat{\rho}_-(s)},
\end{equation}
and
\begin{equation}
\hat{W}_{-+}(s)=\frac{\hat{f}_{E,-}(s)}{s} \frac{1-\hat{\rho}_+(s)}{1 -\hat{\rho}_+(s) \hat{\rho}_-(s)}.
\end{equation}
It follows that the Laplace transform of $C(t)$ is given by
\begin{equation}
\hat{C}(s)= \left(\frac{D_+ - D_- }{\mu_+ + \mu_-}\right)^2 \frac{\mu_+\mu_-}{s}
- \frac{(D_+ - D_-)^2}{\mu_+ + \mu_-} \frac{\{1-\hat{\rho}_+(s)\} \{1-\hat{\rho}_-(s)\}}{\{1-\hat{\rho}_+(s)\hat{\rho}_-(s)\}s^2}.
\end{equation}
\end{widetext}

When both of the duration-time PDFs follow exponential distributions, we have
\begin{equation}
\hat{C}(s) = \frac{(D_+ - D_-)^2\mu_+\mu_-}{(\mu_+ + \mu_-)^2} \frac{1}{(\mu_++\mu_-)/(\mu_+\mu_-) + s}.
\end{equation}
The inverse Laplace transform yields 
\begin{equation}
C(t) = \frac{(D_+ - D_-)^2\mu_+\mu_-}{(\mu_+ + \mu_-)^2} \exp \left(-\frac{\mu_++\mu_-}{\mu_+\mu_-}t\right).
\end{equation}
For $\alpha>2$ (Case 3), we have 
\begin{equation}
\hat{C}(s) = \frac{(D_+ - D_-)^2 (\mu_+\mu_-)^2}{2(\mu_+ + \mu_-)^3} \left(\frac{\sigma_+^2}{\mu_+^2} + \frac{\sigma_-^2}{\mu_-^2}\right) .
\end{equation}
in the small $s \ll 1$.
For $1<\alpha<2$ (Case 2), we have 
\begin{equation}
\hat{C}(s) \sim \frac{(D_+ - D_-)^2 }{(\mu_+ + \mu_-)^3} a_+ \mu_-^2 s^{\alpha-2}
\end{equation}
for $s \to 0$. The inverse Laplace transform reads
\begin{equation}
C(t) \sim \frac{a_+ \mu_-^2(D_+ - D_-)^2 }{\Gamma(2-\alpha)(\mu_+ + \mu_-)^3} t^{-(\alpha-1)}
 \quad (t\to\infty). 
\end{equation}


\section{Application to Langevin equation with alternately fluctuating diffusivity }
Here, we discuss how our results can be applied to stochastic processes such as the Langevin equation with fluctuating diffusivity. 
The continuous-time random walk is described by the Langevin equation with alternating fluctuating diffusivity  \cite{Kimura2022}. In particular, instantaneous 
diffusivity $D(t)$ is given by $D(t)=1$ or 0, which correspond to $\sigma(t)=-1$ or $1$, respectively. In this model, the mean square displacement (MSD) becomes
\begin{equation}
\langle x(t)^2 \rangle = 2 \int_0^t  \langle D(t') \rangle dt' 
\end{equation}
where $x(t)$ is the displacement with $x(0)=0$. When the mean duration time of $-$ state is finite, 
the MSD can be described by the number of changes of states: 
$\langle x(t)^2 \rangle  \sim  \mu_- H(t)$. Therefore,  the MSD becomes $\langle x(t)^2 \rangle  \sim 2 \langle D \rangle_{\rm eq} t$ in Cases 2 and 3 and 
exhibits anomalous diffusion: $\langle x(t)^2 \rangle  \propto t^{\alpha}$ in Case  4, where $\langle D \rangle_{\rm eq}$ is the ensemble average of $D$ in equilibrium and  
defined by Eq.~(\ref{average D eq}). In Cases 2 and 3,  $\langle D \rangle_{\rm eq}$ becomes $\langle D \rangle_{\rm eq}=\mu_-/(\mu_+ + \mu_-)$.

We discuss ergodic properties of the MSD in the Langevin equation with alternating fluctuating diffusivity, i.e., $D(t)=D_+$ or $D_-$ 
\cite{Miyaguchi2016,Akimoto-Yamamoto2016,Miyaguchi2019}. 
The MSD can be defined by the time average:
\begin{equation}
\overline{\delta^2 (\Delta; t)} = \frac{1}{t-\Delta} \int_0^{t-\Delta}  dt' [(x(t'+\Delta) -x(t')]^2.
\end{equation}
If the system is ergodic, the MSD defined by the time average $\overline{\delta^2 (\Delta; t)}$ converges to $\langle x(\Delta)^2 \rangle $ for all $\Delta$ in the long-time 
limit ($t\to \infty$). The time-averaged MSD can be represented by occupation time $T_t^+$:
\begin{equation}
\overline{\delta^2 (\Delta; t)} \sim 2 \left( D_- +  \frac{ (D_+ - D_-)T_t^+}{t} \right) \Delta. 
\end{equation}
Therefore, our results in occupation time statistics of $T_t^+$ can be applied to the time-averaged MSD. For Cases 2 and 3, the ensemble average of 
$\overline{\delta^2 (\Delta; t)}$ converges to $2 \langle D \rangle_{\rm eq} \Delta$. Moreover, the variance of $\overline{\delta^2 (\Delta; t)}$ becomes zero 
for $t\to \infty$ because of the results Eqs.~(\ref{T RSD cases23}) and (\ref{T RSD cases23 eq}). Therefore, the system is ergodic in the sense 
that $\overline{\delta^2 (\Delta; t)} \to  2 \langle D \rangle_{\rm eq} \Delta$ for $t\to \infty$.

\section{Conclusion}

We have investigated the statistics of the number of renewals and the occupation time statistics in alternating renewal processes. 
We analytically obtain the recurrence time distributions for the ordinary alternating renewal process and show that there is an equilibrium distribution when 
the mean duration time exists. On the other hand, when the mean duration time diverges, there is no equilibrium distribution for the recurrence time distribution 
and the system exhibits aging. In other words, the recurrence time distribution explicitly depends on the elapsed time of the system, i.e. aging time $t_a$. 
Therefore, we have considered  the statistics of the number of renewals and the occupation time statistics for ordinary, equilibrium, and 
aging alternating renewal processes. 

Here, we summarize the results of the statistics of the number of renewals. When both of the duration-time PDFs have finite variance (Case 3), the renewal function 
and the variance of the number of renewals increase linearly with for ordinary, equilibrium, and aging alternating renewal processes. 
When one of the duration-time PDFs follow a power-law distribution with time divergent second moment (Case 2), the renewal function increases linearly with time but 
the variance of the number of renewals exhibits a power-law increasing for the ordinary and equilibrium alternating renewal processes. Moreover, the coefficients of 
the variances for the ordinary and equilibrium alternating renewal processes do not coincide in Case 2. 
When the means of duration times diverge (Cases 1 and 4), the renewal function increases sublinearly with time 
and the distribution of the number of renewals converges to the Mittag-Leffer distribution in the long-time limit for the ordinary alternating renewal process.  
In aging alternating renewal processes (Cases 1 and 4), the renewal function depends explicitly on the aging time $t_a$. 

We summarize the results for the occupation time statistics. When one of the duration-time PDFs follow a power-law distribution with time divergent second moment (Case 2), 
the relative standard deviation of occupation times as well as the correlation function exhibit a power-law decay.  
When the means of duration times diverge (Cases 1 and 4), the distribution of the ratio of occupation time follows the generalized arcsine distribution.

\section*{}

{\bf Acknowledgement} T.A. was supported by JSPS Grant-in-Aid for Scien- tific Research (No. C 21K033920).






%

\end{document}